\documentstyle[12pt]{article}
\def\Re{{\cal R \mskip-4mu \lower.1ex \hbox{\it e}\,}}
\def\Im{{\cal I \mskip-5mu \lower.1ex \hbox{\it m}\,}}

\def\etal{{\it et al.}}

\def\sub#1{_{\lower.25ex\hbox{$\scriptstyle#1$}}}
\def\sul#1{_{\kern-.1em#1}}
\def\sll#1{_{\kern-.2em#1}}
\def\sbl#1{_{\kern-.1em\lower.25ex\hbox{$\scriptstyle#1$}}}
\def\ssb#1{_{\lower.25ex\hbox{$\scriptscriptstyle#1$}}}
\def\sbb#1{_{\lower.4ex\hbox{$\scriptstyle#1$}}}

\def\to{\rightarrow}
\def\mh{\ifmmode m\sbl H \else $m\sbl H$\fi}
\def\mch{\ifmmode m_{H^\pm} \else $m_{H^\pm}$\fi}
\def\mt{\ifmmode m_t\else $m_t$\fi}
\def\mc{\ifmmode m_c\else $m_c$\fi}
\def\mz{\ifmmode M_Z\else $M_Z$\fi}
\def\mw{\ifmmode M_W\else $M_W$\fi}
\def\mws{\ifmmode M_W^2 \else $M_W^2$\fi}
\def\mhs{\ifmmode m_H^2 \else $m_H^2$\fi}
\def\mzs{\ifmmode M_Z^2 \else $M_Z^2$\fi}
\def\mts{\ifmmode m_t^2 \else $m_t^2$\fi}
\def\mcs{\ifmmode m_c^2 \else $m_c^2$\fi}
\def\mchs{\ifmmode m_{H^\pm}^2 \else $m_{H^\pm}^2$\fi}
\def\ztwo{\ifmmode Z_2\else $Z_2$\fi}
\def\zone{\ifmmode Z_1\else $Z_1$\fi}
\def\mtwo{\ifmmode M_2\else $M_2$\fi}
\def\mone{\ifmmode M_1\else $M_1$\fi}
\def\tb{\ifmmode \tan\beta \else $\tan\beta$\fi}
\def\xw{\ifmmode x\sub w\else $x\sub w$\fi}
\def\ch{\ifmmode H^\pm \else $H^\pm$\fi}
\def\lum{\ifmmode {\cal L}\else ${\cal L}$\fi}
\def\inpb{\ifmmode {\rm pb}^{-1}\else ${\rm pb}^{-1}$\fi}
\def\infb{\ifmmode {\rm fb}^{-1}\else ${\rm fb}^{-1}$\fi}
\def\epem{\ifmmode e^+e^-\else $e^+e^-$\fi}
\def\ppb{\ifmmode \bar pp\else $\bar pp$\fi}

\newskip\zatskip \zatskip=0pt plus0pt minus0pt
\def\matth{\mathsurround=0pt}

\def\atversim#1#2{\lower0.7ex\vbox{\baselineskip\zatskip\lineskip\zatskip
  \lineskiplimit 0pt\ialign{$\matth#1\hfil##\hfil$\crcr#2\crcr\sim\crcr}}}

\renewcommand{\thefootnote}{\fnsymbol{footnote}}

\begin{document} \begin{titlepage}
\setcounter{page}{1}
\thispagestyle{empty}
\rightline{\vbox{\halign{&#\hfil\cr
&ANL-HEP-PR-92-58\cr
&July 1992\cr}}}
\vspace{0.2in}
\begin{center}

{\Large\bf
The Production of $Z'$ Associated With Photons or Jets as a Probe of New
Gauge Boson Couplings}
\footnote{Research supported by the
U.S. Department of
Energy, Division of High Energy Physics, Contracts W-31-109-ENG-38
and W-7405-Eng-82.}
\medskip

\normalsize THOMAS G. RIZZO
\\ \smallskip
High Energy Physics Division\\Argonne National
Laboratory\\Argonne, IL 60439\\
\smallskip
and\\
\smallskip
Ames Laboratory and Department of Physics\\
Iowa State University\\ Ames , IA 50011\\

\end{center}

\begin{abstract}

We examine the production of a new $Z'$ gauge boson in association with
photons or jets at future hadron supercolliders as a probe of its couplings to
fermions. Associated jet production is found to be rather insensitive
to these couplings and suffers from large uncertainties as well as substantial
backgrounds. On the other hand, the ratio of rates
for associated photon $Z'$ production to that of conventional $Z'$
production has a rather clean signature (once appropriate cuts are made), and
is found to be quite
sensitive to the choice of extended electroweak model, while being
simultaneously insensitive to structure function uncertainties and QCD
corrections. Rates at both
the SSC and LHC are significant for $Z'$ masses in the 1 TeV range.

\end{abstract}



\renewcommand{\thefootnote}{\arabic{footnote}} \end{titlepage}


It is by now well known that the production of a new neutral gauge boson,
$Z'$, in the few TeV mass range should be easily observed at the SSC and LHC
hadron supercolliders {\cite {bigref}} via its decay to lepton pairs. If such
a particle is observed it will be mandatory to determine its couplings to
fermions in order to identify which $Z'$, of the many proposed in the
literature, has been discovered. During the past 1-2 years, this subject has
gotten significant attention from several groups of
authors{\cite {us,lang,taxil,cahn,dela}}
who have found that $Z'$ identification is a serious problem
for realistic detectors if as few as possible theoretical assumptions are made
about the $Z'$ decay modes even if such new gauge bosons are relatively light.
{\it If} one assumes that the $Z'$ can decay only to the conventional
particles of the Standard Model(SM) then it has been shown{\cite {us}} that
measurements of its mass($M_2$), width($\Gamma_2$), and production cross
section($\sigma_0$), together with the corresponding leptonic forward-backward
asymmetry($A_{FB}^l$), can be used to `identify' the $Z'$ for masses up to
several TeV. However, we note that many extended electroweak models(EEM)
allow for non-SM $Z'$ decays which could dominate the $Z'$ width although the
above assumption is not so bad in some specific cases. Of the
observables listed above,
only $A_{FB}^l$ (other than, of course, $M_2$) is insensitive to any
assumptions about the $Z'$ decay modes and so, by itself, is insufficient to
probe the details of the new gauge boson's couplings. It is thus absolutely
necessary to find additional observables which are also insensitive to any
assumption on how the $Z'$ may decay.

One suggestion {\cite {combo}} is to look for multi-body $Z'$ fermionic decay
modes and to form various ratios of decay rates and a second is to examine the
polarization of $\tau$'s resulting from the decay $Z'\to \tau^+\tau^-$
{\cite {cahn}}. A third proposal takes advantage of the potential possibility
of {\it polarized} pp scattering {\cite {taxil}} to create a sizeable left-
right asymmetry. All of these scenarios suffer from either large SM
backgrounds which must be subtracted (but are still found to be useful for
a relatively
light $Z'$ of order 1 TeV in mass) or are hampered by our current lack of
knowledge of the polarized parton distributions.

Recently, Cvetic and Langacker{\cite {lang2}} have proposed the use of
associated $Z'$ production, i.e., ${\bar q}q\to VZ'$, with $V=Z,W^{\pm}$,
as a new probe of the $Z'$ couplings to fermions. The ratios of the
cross sections for these events to that for single $Z'$ production (as
measured via the $Z'\to l^+l^-$ channel) are independent of $\Gamma_2$,
were found to be statistically significant in the absence of cuts, and quite
sensitive to the choice of EEM. Of course, paying the price of applying
realistic cuts
and allowing for $V$ branching fractions (or $V$ identification efficiencies)
will reduce the values of these ratios somewhat resulting in a significant
decrease in model sensitivity via a loss is statistical power.

In this paper we will examine both $Z'$ produced together with a single jet
or together with an isolated photon; the first process proceeds in lowest
order{\cite {halz}} either via ${\bar q}q\to Z'g$ or $gq\to Z'q$ while the
second proceeds only via ${\bar q}q\to Z'\gamma$ {\cite {old}} in lowest
order. Although the $gq$ production process was ignored in the brief
discussion given by Cvetic and Langacker, we verify their conclusion that $Z'$
production in association with a jet is quite insensitive to the $Z'$
couplings to fermions. $Z'\gamma$ production, on the other hand, will be shown
to be very clean and effectively background free when only very mild cuts
are applied. Additionally, the efficiency of isolated photon detection is
very high for planned collider detectors{\cite {detect}} due to its usefulness
in hunting for the intermediate-mass Higgs boson of the SM. We will show
below that the ratio of the number of $Z'\gamma$ to $Z'$ events observed at
either the SSC or LHC, detected via the leptonic decay of the $Z'$, provides
a statistically useful probe of the $Z'$ couplings which is insensitive to
variations in the parton densities and higher order QCD corrections. Unlike
the situation of $Z'V$ production, in the $Z'\gamma$ case we need not pay any
significant price in applying cuts to remove SM backgrounds or for $V$
branching fractions.

There are very many models in the literature which predict the existence of a
$Z'$ so that we can hardly perform an exhaustive analysis. Thus to be
specific we'll deal with only a small representative set of EEM's which we
feel are fairly representative: (i) the `Effective Rank-5' Models (ER5M)
arise from string-inspired $E_6$ {\cite {er5m}} and are obtained via the
symmetry breaking chain
$E_6\to SO(10)\times U(1)_\psi \to SU(5)\times U(1)_\chi \times U(1)_\psi
\to SM\times U(1)_\theta$ such that we can identify $Z'=Z_\psi cos\theta-Z_\chi
sin\theta$ with $-\pi /2\leq \theta \leq \pi/ 2$ being an a priori
free parameter
whose value fixes the $Z'$ couplings to fermions; (ii) the now-classic Left
Right Model(LRM){\cite {lrm}} with $g_L=g_R$; (iii) the `Alternative' Left
Right Model(ALRM)
{\cite {alrm}}; (iv) a toy model wherein the $Z'$ is just a heavier version of
the SM Z (SSM). We refer the reader to the original literature for the details
on each of these EEM's.

Following Ref.9, the lowest order $Z'+jet$ or $Z'\gamma$ differential
production cross section can be written as
\medskip
\begin{equation}
{d\sigma \over dp_tdy} = 2p_t{\sum_{ij}}{\int^1_{x_{min}}}{{\hat
{s}}f_i(x_1,q^2
)
f_j(x_2,q^2){\hat {\sigma}}_{ij}(\hat {s},\hat {t},\hat {u})\over {x_1s+u
-M_2^2}}
\end{equation}
\medskip
The kinematics are defined via the relationships
\medskip
\begin{eqnarray}
m_T^2 &=& p_t^2 + M_2^2 \nonumber \\
\hat {s} &=& sx_1 x_2 \nonumber \\
t,u &=& -{\sqrt s}~m_T~e^{\mp y}+M_2^2 \nonumber \\
\hat {t},\hat {u} &=&  -{\sqrt s}~m_T~x_{1,2}~e^{\mp y}+M_2^2 \\
x_2 &=& -x_1 t -(1-x_1)M_2^2\over {x_1s+u-M_2^2} \nonumber \\
x_{min} &=& -u\over {s+t-M_2^2} \nonumber
\end{eqnarray}
\medskip
and $f_i$ are the appropriate parton densities. For ${\bar q}q\to Z'g$ we have
\medskip
\begin{equation}
{\hat {\sigma}}_{{\bar q}q} = {2{\sqrt 2}G_FM_Z^2\over {9{\hat {s}}^2}}
\biggl({{\hat {t}}\over {\hat {u}}} +
{{\hat {u}}\over {\hat {t}}} +{2{\hat {s}}M_2^2
\over {\hat {u}}{\hat {t}}}\biggr) \alpha_s(q^2)(v_i^2+a_i^2)
\end{equation}
\medskip
whereas for ${\bar q}q\to Z'\gamma$, we replace $\alpha_s(q^2)$ by
$3/4~\alpha(q^2)Q_i^2$ where $Q_i$ is the quark electric charge in units of
e. For the $gq\to Z'q$ subprocess one has instead
\medskip
\begin{equation}
{\hat {\sigma}}_{gq} = {{\sqrt 2}G_FM_Z^2\over {12{\hat {s}}^2}} \biggl
(-{{\hat {s}}\over {\hat {u}}} - {{\hat {u}}\over {\hat {s}}} -{2{\hat
{t}}M_2^2
\over {{\hat {u}}{\hat {s}}}}\biggr) \alpha_s(q^2)(v_i^2+a_i^2)
\end{equation}
\medskip
In writing down these expressions we have normalized the various fermionic
couplings to the $Z'$ as in the SM:
\medskip
\begin{equation}
{\cal L} = {g\over {2c_w}} {\bar q}_i{\gamma_\mu}(v_i-a_i\gamma_5)q_iZ'_\mu
\end{equation}
\medskip
with $c_w=cos\theta_w$ and $g$ being the usual weak coupling constant. For
purposes of numerical evaluations we take $q^2=M^2_2$ and evolve
$\alpha_s(q^2)$ via the 3-loop renormalization group equation (taking the
appropriate value of the scale $\Lambda$ associated with the choice of
parton distributions); we also take $\alpha^{-1}(q^2)$=127.9.

Let us first briefly examine the $Z'$ plus jet production process; we
normalize our differential rates by the lowest order ${\bar q}q\to Z'$
production cross section, $\sigma_0$. Since the Q=2/3 and Q=-1/3 quarks
contribute differently to the two distinct subprocesses one might expect
that the $Z'$ plus jets production rate might be sensitive to the fermionic
$Z'$ couplings; unfortunately this is not the case. Fig. 1a shows
the normalized
differential rate for the SSC as a function of the jet $p_t$ for $y=0$
assuming the Morfin-Tung set S1 (MT-S1) parton distributions{\cite {mt}}
taking $M_2$ = 1 TeV for four different EEM's. Although this is only a Born
level calculation, we see the essential feature immediately: all of the
predictions lie virtually atop one another over a wide range of $p_t$.
Fixing the $p_t$ at 300 GeV and maintaining $y=0$, Fig. 1b shows the extremely
weak $\theta$ dependence (about $10\%$) of the normalized $Z'$ plus jet cross
section for the ER5M which again demonstrates the lack of sensitivity of this
mechanism to the fermionic $Z'$ couplings anticipated by the discussion given
by Cvetic and Langacker{\cite {lang2}}. We thus conclude that this reaction is
useless as a probe of the $Z'$ couplings. We note, however, that had the color
factors been such as to make the gq subprocess occur at an even larger rate
then the $Z'$ plus jet mode might have provided a relatively sensitive tool
with which to have analyzed the $Z'$ couplings.

Turning now to the $Z'\gamma$ mode we see in Fig. 2a the normalized
differential rate for this process as a function of the photon's $E_t$ for
the same situation as in Fig. 1a. Instead of lying atop one another, we see
here that the predictions of the four different EEM's yield somewhat different
results giving us some hope of the usefulness of this channel. Of course,
since the rates are small and differential distributions are more sensitive to
QCD corrections than are integrated quantities, we integrate our distribution
over the photon $E_t>$ 50 GeV and the rapidity interval
\medskip
\begin{equation}
|y|~ \leq~ min\biggl[2.5, cosh^{-1}\biggl(
{s+M_2^2\over {2{\sqrt s}m_T}}\biggr)\biggr]
\end{equation}
\medskip
Here the former value represents the typical $\gamma$ rapidity coverage of
the SSC and LHC detectors{\cite {detect}} while the latter is purely kinematic.
(A similar rapidity cut can be applied to the leptons from the decay of the
$Z'$.) Backgrounds from decays such as $Z'\to l^+l^-\gamma$ can be
completely removed by demanding that the lepton pair mass satisfy
$M_{ll}>0.95M_2$ coupled with the photon's $E_t$ cut for a $Z'$ with a
mass of 1 TeV. Note that the typical supercollider detector will have a
dilepton pair mass resolution of order $1\%$ or better{\cite {detect}}. As long
as the probability of mis-identifying a jet as a photon is less than about
$10^{-3}$, there are no significant backgrounds from QCD sources which are not
removed by the above cuts. This level of jet rejection should be obtainable
for most of the SSC and LHC detectors{\cite {detect}}.

The ratio of $Z'\gamma$ to $Z'$ events, $R_\gamma$, is shown for the SSC
assuming $M_2$ = 1 TeV for the ER5M case as a function of the parameter
$\theta$

in Fig. 2b for several different choices of the parton
densities{\cite {mt,hmrs,kmrs}}.
Here we see that (i) the results are insensitive to the choice of
parton densities with a variation of at most $5\%$ for the models we've
examined; (ii) $R_\gamma$ lies in the range 0.2-0.9$\%$; and (iii) $R_\gamma$
is quite sensitive to the value of $\theta$ as we would hope. Assuming MT-S1
distributions we also find that $R_\gamma$=(4.95, 8.46, 5.50)$10^{-3}$
corresponding to the (LRM, ALRM, SSM) cases respectively. For the LHC, under
identical assumptions for the same models we find instead that $R_\gamma$=
(4.65, 7.26, 5.11)$10^{-3}$, numerically comparable to their corresponding
values at the SSC. For the ER5M case, the predicted value of $R_\gamma$ at
the LHC is shown in Fig. 2c as a function of $\theta$ assuming the same sets
of structure functions as in Fig. 2b.

For larger $Z'$ masses, e.g., $M_2$ = 3 TeV, the ratio $R_\gamma$ is somewhat
increased as shown in Fig. 2d and has a comparable sensitivity to variations
in the $Z'$ couplings. In fact, $R_\gamma$ is found to approximately scale
with the $Z'$
mass and choice of minimum photon $E_t$ as $log^2({M_2}/{E_t^{min}})$.
However, since the {\it number} of $Z'$ events is drastically smaller for
the larger $Z'$ mass we lose the statistical power of $R_\gamma$ as will be
apparent from the number of events that we present below in the case of $M_2$ =
1 TeV.

Since we have so far presented only a Born-level calculation, we must worry
about how $R_\gamma$ would be modified by QCD corrections; such corrections
have been considered in the literature for the production of $Z\gamma$ and
$W^{\pm}\gamma${\cite {box}}. One possibly sizeable correction at SSC and
LHC energies
arises from the box diagram-mediated process $gg\to Z'\gamma$. In the SM case,
this represents an approximate $30\%$ effect due to the high $gg$ luminosity at
small x values. In the $Z'\gamma$ case this contribution will be much smaller
as significantly larger x values are being probed since the $Z'$ is so
massive. Additionally, this contribution is {\it model dependent} as it is
sensitive to the existence of all color non-singlet fields in the model which
couple to the $Z'$ and the photon.  Full next-to-leading(NLL) order
calculations of $Z\gamma$ production in pp collisions have only recently been
completed by Ohnemus{\cite {box}}; we note that the choice of kinematic cuts
selected by that author is quoting his results is identical to the choice we
have made above. Thus we can estimate that the corrections to the integrated
$Z'\gamma$ cross section at both the SSC and LHC will be almost identical to
the size of the `K-factor' correction to the total $Z'$ production rate as
given, e.g., by the analysis of Hamberg et al.{\cite {kfact}} which we have
used in our earlier work{\cite {us}}. This being the case, we estimate that the
numerical values of $R_\gamma$ presented above are relatively insensitive to
large higher order QCD corrections at the level of more than a few percent. In
quoting the numbers of events below,
we will take all such `K-factor' effects into account.

How well can $R_\gamma$ be determined? Since there is little background and
many of the various systematic uncertainties cancel in forming the ratio of
cross sections, the dominant error in $R_\gamma$ is expected to be statistical
so that it will scale  approximately inversely proportional to the square
root of
the number of $l^+l^-\gamma$ events($N_\gamma$) which pass our cuts. We will
assume that
the isolated lepton identification efficiency is 0.85 separately
for {\it both} e's and
$\mu$'s and will sum over both leptonic flavors below. Table 1 shows the
resulting values of $N_\gamma$ for both the SSC (L=10$fb^{-1}$) and LHC
(L=100$fb^{-1}$) with $M_2$ = 1 TeV and assuming MT-S1 parton distributions for
several different EEM. The
Table also shows the anticipated size of the relative error on a $R_\gamma$
measurement for each of the EEM at both colliders. With the integrated
luminosities that we've assumed, it is clear that $R_\gamma$ can be relatively
well determined at either supercollider for a 1 TeV $Z'$ although the
anticipated errors for the LHC are somewhat smaller due to the approximate
factor of 2 larger event rate. It is important to note that the assumed factor
of 10 larger luminosity of the LHC {\it only} translates into an approximate
factor of 2 larger rate due to the LHC's smaller center of mass energy. It is
clear from the numbers in the Table that this method will fail for $Z'$ masses
significantly larger than 1 TeV since the event rates will fall off quite
rapidly with increasing $Z'$ mass. Thus this technique is seen to be limited
to the case of a relatively light $Z'$.

In this paper we have obtained the following results:
\begin{itemize}
\item[{\it (i)}] By explicit calculation, we demonstrated that the associated
production of $Z'$ plus jets is insensitive to the fermionic couplings of the
$Z'$ even though two distinct subprocesses contribute to the full cross
section.
\item[{\it (ii)}] We have shown that the ratio of the cross sections for
$Z'\gamma$ and $Z'$ production, $R_\gamma$, is a sensitive probe of the $Z'$
couplings, and is insensitive to structure function uncertainties and QCD
corrections when suitable `K-factor' contributions are accounted for.
\item[{\it (iii)}] With suitably soft cuts which do not modify the signal rate,
$Z'\gamma$ production is found to be essentially free of QCD and radiative
$Z'$ decay backgrounds with a final state that can be easily identified with
high efficiency without paying the price of small branching fractions.
\item[{\it (iv)}] Although sufficient statistics can be accumulated at both the
SSC and LHC to make $R_\gamma$ a useful tool for a 1 TeV $Z'$, the
event rate falls off quite quickly with increasing mass rendering it useless
if the $Z'$ is significantly heavier.
\end{itemize}
Hopefully a new $Z'$ will exist in the mass range of interest and provide us
with further clues to new physics beyond the SM.

\newpage
\vskip.25in
\centerline{ACKNOWLEDGEMENTS}

The author with like to thank J. Hewett and J. Ohnemus for discussions
related to this work.
This research was supported in by the U.S.~Department of Energy under
contracts W-31-109-ENG-38 and W-7405-ENG-82.

\newpage

%
\def\MPL #1 #2 #3 {Mod.~Phys.~Lett.~{\bf#1},\ #2 (#3)}
\def\NPB #1 #2 #3 {Nucl.~Phys.~{\bf#1},\ #2 (#3)}
\def\PLB #1 #2 #3 {Phys.~Lett.~{\bf#1},\ #2 (#3)}
\def\PR #1 #2 #3 {Phys.~Rep.~{\bf#1},\ #2 (#3)}
\def\PRD #1 #2 #3 {Phys.~Rev.~{\bf#1},\ #2 (#3)}
\def\PRL #1 #2 #3 {Phys.~Rev.~Lett.~{\bf#1},\ #2 (#3)}
\def\RMP #1 #2 #3 {Rev.~Mod.~Phys.~{\bf#1},\ #2 (#3)}
\def\ZP #1 #2 #3 {Z.~Phys.~{\bf#1},\ #2 (#3)}
\def\IJMP #1 #2 #3 {Int.~J.~Mod.~Phys.~{\bf#1},\ #2 (#3)}

\newpage

\begin{table}
\caption{The number of $Z'\gamma$ events $(N_\gamma)$ and the relative error
in $R_\gamma$ in percent for several EEM's at both the SSC and LHC assuming
MT-S1 parton distributions.}
\vspace{1.0in}
\begin{center}
\begin{tabular}{|c|c|c|} \hline \hline
EEM     & $N_\gamma$     & $\delta R_\gamma/R_\gamma~(\%)$  \\ \hline
\multicolumn{3}{|c|}{SSC (10 fb$^{-1}$)} \\ \hline
LRM   & 65.4   & 12.4 \\
ALRM  & 180.7  & 7.4  \\
SSM   & 109.6  & 9.6  \\
$\psi$& 26.8   & 19.3 \\
$\chi$& 40.0   & 15.8 \\
$\eta$& 39.0   & 16.0 \\ \hline
\multicolumn{3}{|c|}{LHC (100 fb$^{-1}$)} \\ \hline
LRM   & 125.2 & 8.9  \\
ALRM  & 393.6 & 5.0  \\
SSM   & 207.5 & 6.9  \\
$\psi$& 63.4  & 12.6 \\
$\chi$& 74.6  & 11.6 \\
$\eta$& 81.7  & 11.0 \\ \hline
\end{tabular}
\end{center}
\vspace{1.0in}
\end{table}

\pagebreak
%
{\bf Figure Captions}
\begin{itemize}

\item[Figure 1.]{(a) Normalized Born-level $p_t$ distribution for $Z'$ plus
jet production at the SSC with $y=0$ assuming $M_2$ = 1 TeV and MT-S1 parton
distributions. The solid(dash-dotted, dashed, dotted)curve corresponds to
the LRM($\chi$, $\psi$, ALRM) case. (b) Same as (a) but for the ER5M as a
function of $\theta$ assuming $p_t$= 300 GeV}
\item[Figure 2.]{(a) Normalized Born-level $E_t$ distribution for $Z'\gamma$
production at the SSC with $y=0$ assuming $M_2$ = 1 TeV and MT-S1 parton
distributions. The solid(dash-dotted, dashed, dotted) curve corresponds to
the LRM(ALRM, $\psi$, $\chi$) case. (b) The ratio $R_\gamma$ assuming a 1 TeV
$Z'$ at the SSC after cuts for the ER5M as a function of $\theta$. The solid(
dash-dotted, dashed, dotted, square dotted) curve corresponds to  the choice of
MT-S1(HMRS-B, MT-S2, KMRSB0, KMRSB-2) parton densities. (c) Same as (b) but
for the LHC assuming the same sets of  parton distributions. (d) Same as (b)
but for a 3 TeV $Z'$ at the SSC.}

\end{itemize}


\begin{thebibliography}{99}
\bibitem{combo}
This has been discussed by the authors of Refs. 2, 3, and 6.
\end{thebibliography}
\end{document}